\documentclass[aps,twocolumn,pra,superscriptaddress,showkeys,floatfix]{revtex4-1}

\usepackage{amsbsy}
\usepackage[pdftex]{graphicx}
\usepackage{amsmath}
\usepackage{amssymb}
\usepackage{amsfonts}
\usepackage{amssymb}
\usepackage{float}
\usepackage{graphicx}
\usepackage{enumerate}
\usepackage{epstopdf}
\usepackage{braket}
\usepackage{bm}
\usepackage{color,soul}

\begin{document}

\title{Applications of Picard and Magnus expansions to the Rabi model}

\author{Fabrizio Angaroni}
\affiliation{Center for Nonlinear and Complex Systems,
Dipartimento di Scienza e Alta Tecnologia,
Universit\`a degli Studi dell'Insubria, via Valleggio 11, 22100 Como, Italy}
\affiliation{Istituto Nazionale di Fisica Nucleare, Sezione di Milano,
via Celoria 16, 20133 Milano, Italy}
\author{Giuliano Benenti}
\affiliation{Center for Nonlinear and Complex Systems,
Dipartimento di Scienza e Alta Tecnologia,
Universit\`a degli Studi dell'Insubria, via Valleggio 11, 22100 Como, Italy}
\affiliation{Istituto Nazionale di Fisica Nucleare, Sezione di Milano,
via Celoria 16, 20133 Milano, Italy}
\affiliation{NEST, Istituto Nanoscienze-CNR, 56126 Pisa, Italy}
\author{Giuliano Strini}
\affiliation{Department of Physics, University of Milan,
via Celoria 16, 20133 Milano, Italy}

\begin{abstract}
We apply the Picard and Magnus expansions to both the semiclassical and the 
quantum Rabi model, with a switchable matter-field coupling. 
The case of the quantum Rabi model ia a paradigmatic example of 
finite-time quantum electrodynamics (QED), and in this case we build an
intuitive diagrammatic representation of the Picard series.  
In particular, we show that regular oscillations in the mean number of 
photons, ascribed to the dynamical Casimir effect (DCE) 
for the the generation of photons and to the anti-DCE for their 
destruction, take place at twice the resonator frequency 
$\omega$. Such oscillations, which are a clear dynamical 
``smoking gun'' of the ultrastrong
coupling regime,  can be predicted by first-order Picard expansion.  
We also show that the Magnus expansion can be used, through concatenation,  
as an efficient numerical integrator for both the semiclassical and the 
quantum Rabi model. In the first case, we find distinctive features in
the Fourier spectrum of motion, with a single peak at the Rabi frequency
$\Omega$ and doublets at frequencies $2n\omega\pm\Omega$, 
with $n$ positive integer. 
We explain these doublets, which are a feature beyond 
the rotating wave approximation (RWA), on the basis of the Picard series. 
\end{abstract}


\maketitle

\section{Introduction}

Ultrastrong coupling between artificial atoms and electromagnetic 
cavity modes is achieved when the coupling strength
$\lambda$ becomes comparable to, or even exceeds
the resonator frequency $\omega$. Such regime, which is 
nowadays experimentally addressed in circuit 
QED~\cite{bourassa,gross,mooij,lupascu,semba}, 
is of interest both for the development of quantum technologies and 
for fundamental physics. Indeed, strong matter-field coupling 
is preliminary to the implementation of fast quantum protocols. 
On the other hand, in the ultrastrong coupling regime strongly 
correlated matter-light states emerge~\cite{lupascu,semba}.

A prominent phenomenon in ultrastrong matter-field coupling 
is the dynamical Casimir effect, that is, the generation of photons 
from the vacuum due to time-dependent boundary conditions or, more generally, 
as a consequence of the \emph{nonadiabatic} change of some parameters of a 
system \cite{moore,dodonov,noriRMP} 
(this latter case is usually refereed to as parametric DCE \cite{dodonov}). 
DCE has been discussed in several contexts, for instance
in Bose-Einstein condensates~\cite{jaskula}, in excition-polariton
condensates~\cite{koghee}, for multipartite entanglement generation
in cavity networks~\cite{solano2014}, in relation to several forms
of quantum correlations~\cite{johansson2015,adesso2015,savasta2015},
in the generation of 
exotic field states~\cite{exotic}§, for quantum
communication protocols~\cite{casimirqip}, in quantum
thermodynamics~\cite{frigo}; the DCE can also be amplified  
via optimal control techniques~\cite{DCEoptimal}. 
Moreover, pioneering experimental demonstrations 
of the DCE have been reported in superconducting circuit 
QED~\cite{norinature,lahteenmaki}. 

In contrast with standard QED, here we consider a single (cavity) mode
rather than an infinite number of modes. Moreover, the quantization 
volume (of the cavity) is fixed and the limit of infinite volume is
not taken at the end. Finally, the interaction is not switched on and
off adiabatically, but we rather focus on \emph{transient} phenomena associated
with the nonadiabatic switching of the matter-field coupling.  
That is, we are considering \emph{finite-time} QED, a problem barely 
considered in the literature~\cite{nomoto}.

The quantum Rabi model~\cite{micromaser,QRM}, 
which describes the dipolar light-matter coupling,
with the addition of a switchable coupling, is the
ideal testing ground to explore finite-time QED in 
the ultrastrong coupling regime.

In this paper, we examine applications of the the Picard and Magnus 
expansions to both the semiclassical and the quantum Rabi model, 
with a time-dependent coupling. While the dynamics of
these models can be addressed 
numerically via a Runge-Kutta integration of the equations of motion,
perturbative methods can shed light on the physical mechanisms
and elementary processes which govern the dynamics. 
We first investigate the Picard series, which allows an intuitive 
diagrammatic representation. Such series, truncated to low orders,
provides a rather accurate description only for 
short interaction times (and not too strong coupling).
In particular, we show that regular oscillations in the 
mean number of photons, can be ascribed to the 
coherent generation (DCE) and destruction 
(anti-DCE~\cite{antiDCE,motazedifard}) of photons. 
Such oscillations take place at a frequency $2\omega$ that can be 
predicted by first-order Picard expansion, and are a clear 
dynamical ``smoking gun'' of the ultrastrong coupling regime. 
We then examine the Magnus expansion, and show that 
through concatenation it can be used as an efficient numerical integrator.
In particular, we study the Fourier spectrum of motion 
for the semiclassical Rabi model and show that it has a 
characteristic structure, with a single peak at the Rabi frequency $\Omega$
and doublets at frequencies $2n\omega\pm\Omega$, with $n=1,2,3,...$.
The doublets, which are a feature beyond the 
RWA, are explained on the basis 
of the Picard series. Finally, 
we discuss analogies between the semiclassical Rabi model and the 
Mathieu equation. 

\section{The finite-time Rabi model}

We consider both the semiclassical and the quantum 
finite-time Rabi models,
describing the interaction of a two-level atom (qubit) with
the electromagnetic field~\cite{micromaser}. In both cases, the Hamiltonian
\begin{equation}
H(t)=H_0+H_I(t),
\end{equation}
where $H_0$ refers to the free evolution for the qubit and the field,
and $H_I(t)$ describes a time-modulated qubit-field coupling,
which extends over a finite time $0\le t\le \tau$.

In the semiclassical Rabi model, which describes within 
the dipole approximation,
the interaction between the qubit and a classical monochromatic field,
(hereafter we set the reduced Planck's constant $\hbar=1$),
\begin{equation}
\begin{array}{c}
{\displaystyle
H_0=-\frac{1}{2}\,\omega_q\sigma_z,
}
\\
\\
{\displaystyle
H_I(t)=f(t)\,[2\Omega \cos(\omega t+\phi)]\,\sigma_x,
}
\end{array}
\end{equation}
where $\omega_q$ and $\omega$ are the qubit and field frequency,
respectively, $\Omega$ is the (Rabi) frequency of the field-induced 
oscillations between the two levels $|g\rangle$ and $|e\rangle$,
the Pauli matrices $\sigma_k$ ($k=x,y,z$) are written in
the $\{|g\rangle,|e\rangle\}$ basis, and the function $f(t)$ 
 modulates the qubit-field coupling. Hereafter, for simplicity's
sake we shall assume 
the phase $\phi= 0$, the resonant case $\omega_q=\omega$,
and a sudden switch on/off of the coupling: $f(t)=1$ for
$0\le t\le \tau$, $f(t)=0$ othertwise.  

In the case of
the quantum Rabi model, which describes the interaction between the qubit 
and a single mode of the quantized field,  
\begin{equation}
\begin{array}{c}
{\displaystyle
H_0=-\frac{1}{2}\,\omega_{q} \sigma_z +
\omega\left(a^\dagger a +\frac{1}{2}\right),
}
\\
\\
{\displaystyle
H_I(t)=f(t)\,[\lambda \,\sigma_+\,(a^\dagger+a)
+\lambda^\star \sigma_-\,(a^\dagger+a)],
}
\end{array}
\label{eq:noRWAparam}
\end{equation}
where
$\sigma_\pm = \frac{1}{2}\,(\sigma_x\mp i \sigma_y)$
are the raising and lowering operators for the qubit
(so that $\sigma_+=|e\rangle\langle g|$ and
$\sigma_-=|g\rangle\langle e|$):
$\sigma_+ |g\rangle = |e\rangle$,
$\sigma_+ |e\rangle = 0$,
$\sigma_- |g\rangle = 0$,
$\sigma_- |e\rangle = |g\rangle$.
The operators $a^\dagger$ and $a$ for the field create
and annihilate a photon:
$a^\dagger |n\rangle=\sqrt{n+1}|n+1\rangle$,
$a |n\rangle=\sqrt{n}|n-1\rangle$,
$|n\rangle$ being the Fock state with $n$ photons.
For the sake of simplicity, from now on we consider a real 
coupling strength, $\lambda\in\mathbb{R}$,
$\omega_q=\omega$ and a 
time-dependent modulation set as above for the semiclassical model. 

The rotating wave approximation  
(valid for $\lambda\to0$)
is obtained neglecting the term
$\sigma_+ a^\dagger$, which simultaneously
excites the qubit and creates a photon,
and $\sigma_- a$, which de-excites the
qubit and annihilates a photon. In this limit, the Hamiltonian
(\ref{eq:noRWAparam}) reduces to the Jaynes-Cummings
Hamiltonian \cite{micromaser} with a time-dependent
modulation.
In the RWA the swapping time needed to transfer
an excitation from the qubit to the field or vice versa
($|e\rangle |0\rangle\leftrightarrow |g\rangle |1\rangle$)
is $\tau_s=\pi/2\lambda$, {and no DCE is possible since the 
total number of excitations in the system is conserved}. 
Within RWA, the (Rabi) frequency of the Rabi oscillations
between the states $|e\rangle |n-1\rangle$ and 
$|g\rangle |n\rangle$ is $\Omega_n=\lambda\sqrt{n}$.

In the interaction picture, the Hamiltonian reads
$\tilde{H}_I(t)=U^\dagger(t)H_I(t) U(t)$,
where $U(t)=e^{-iH_0t}$.
From now on we shall omit tildes and always refer to the interaction 
picture. 
For the semiclassical Rabi model, 
\begin{equation}
H_I(t)=\Omega f(t)\left[(1+e^{-2i\omega t}) \sigma_- + (1+e^{2i\omega t}) \sigma_+\right],
\label{eq:HIsemi}
\end{equation}
while in the quantum Rabi model
\begin{equation}
H_I(t)=\lambda f(t)\,[\sigma_- ae^{-2i\omega t}+\sigma_+a+
\sigma_- a^\dagger 
+\sigma_+ a^\dagger e^{2i\omega t}]. 
\label{eq:HIquantum}
\end{equation} 
In both cases, the RWA is recovered if we neglect the counter-rotating 
terms at frequency $2\omega$.

\section{Picard series}

The solution to the time-dependent Schr\"odinger equation 
$i\,|\dot{\psi}(t)\rangle=H_I(t)\,|\psi(t)\rangle$
can be approximated by the Picard iterative process. We start by writing
the integral associated equation
\begin{equation}
|\psi(t)\rangle=|\psi(0)\rangle-i\int_0^t H_I(t') \,|\psi(t')\rangle \,dt'.
\end{equation}
Iterating the process we obtain
\begin{eqnarray}
\begin{array}{c}
{\displaystyle
|\psi(t)\rangle=|\psi(0)\rangle-i\int_0^{t'} H_I(t')
\left[|\psi(0)\rangle\right.
}
\\
{\displaystyle
\left. -i\int_0^{t'} H_I(t'')\,|\psi(t'')\rangle \,dt''\right]dt',
}
\end{array}
\end{eqnarray}
and so on. Hence we can write 
\begin{equation}
|\psi(t)\rangle=\sum_{k=0}^\infty |\psi^{(n)}(t)\rangle,
\end{equation}
with the zeroth-order approximation
$|\psi^{(0)}(t)\rangle=|\psi(0)\rangle$, the first-order correction
\begin{equation}
|\psi^{(1)}(t)\rangle=-i\int_0^t H_I(t') \,|\psi^{(0)}(t')\rangle \,dt',
\label{eq:firstorder}
\end{equation}
and so on, with the $n$-th-order correction given by 
\begin{equation}
|\psi^{(n)}(t)\rangle=-i\int_0^t H_I(t') \,|\psi^{(n-1)}(t')\rangle \,dt'.
\label{eq:nthorder}
\end{equation}

\subsection{Semiclassical Rabi model}

We expand the state vector in the $\{|g\rangle,|e\rangle\}$ basis for the 
qubit: $|\psi(t)\rangle=C_g(t)|g\rangle+C_e(t)|e\rangle$.
For concreteness, we consider the initial state $|\psi(0)\rangle=|g\rangle$
(however, the considerations of this subsection would not change for
a different initial state).

It is instructive to consider first the RWA limit, in which we easily
obtain the exact solution to the Schr\"odinger equation,
$|\psi(t)\rangle = \cos(\Omega t)|g\rangle-i \sin(\Omega t)|e\rangle$,
corresponding to Rabi oscillations between the two states $|g\rangle$ and 
$|e\rangle$.
In this case, the $n$-th order Picard expansion of 
$|\psi(t)\rangle$ coincides with the
result obtained from the $n$-th order Taylor expansion of the exact 
coefficients 
$C_g(t)=\cos(\Omega t)$ and $C_e(t)=-i \sin(\Omega t)$:
\begin{eqnarray}
\begin{array}{c}
{\displaystyle
|\psi^{(0)}(t)\rangle=|g\rangle, \;
|\psi^{(1)}(t)\rangle=-i (\Omega t)|e\rangle, 
}
\\
{\displaystyle
|\psi^{(2)}(t)\rangle=-\frac{(\Omega t)^2}{2!}|g\rangle, \;
|\psi^{(3)}(t)\rangle=i \frac{(\Omega t)^3}{3!}|e\rangle,...\,. 
}
\end{array}
\end{eqnarray}

Including the counter-rotating terms, we obtain
\begin{eqnarray}
\begin{array}{c}
{\displaystyle
|\psi^{(0)}(t)\rangle=|g\rangle, 
}
\\
{\displaystyle
|\psi^{(1)}(t)\rangle=\left[\frac{\Omega}{2  \omega}\left (1 - 
e^{2 i \omega  t}  \right)-
i \, (\Omega \, t)
\right]  | e \rangle, 
}
\\
{\displaystyle
|\psi^{(2)}(t)\rangle=\left[ \frac{\Omega^{2}}{4  \omega^{2}}\left (-1 
+ e^{2i \omega  t }\right)\right.
}
\\
{\displaystyle
\left.
- i  \frac{ \Omega}{2  \omega} \, (\Omega  t) 
e^ {- 2 i \omega  t }
- \frac{(\Omega  t)^{2}}{2}
\right]  | g \rangle, 
}
\\
{\displaystyle
|\psi^{(3)}(t)\rangle=
\left[ \frac{\Omega^{3}}{8  \omega^{3}} \left(
\frac{5}{2}  -  e^{ - 2 i \omega  t } - 
e^{2 i \omega  t }  -  \frac{1}{2} \, 
e^{4 i \omega  t } \right)\right.
}
\\
{\displaystyle
+\frac{\Omega^{2}}{4  \omega^{2}} \,(\Omega  t) 
\left(1 - e^{ - 2 i  \omega t } + e^{ 2 i  \omega t }\right)
}
\\
{\displaystyle
\left. -\frac{\Omega}{4  \omega} \,(\Omega  t)^{2} \,
\left(
1 -  e^{ 2 i \omega  t }\right) 
+i  \frac{(\Omega  t)^{3}}{6}\right] | e \rangle,...\,.
}
\end{array}
\end{eqnarray}
From these expressions, it is clear that besides the RWA terms 
(Taylor expansions of $\cos(\Omega t)$ and $\sin(\Omega t)$),
we have terms proportional to $e^{n(2i\omega t)}$, multiplied by
powers of $\Omega t$. We will discuss in Sec~\ref{sec:Maghnussemi}
the signatures of these terms in the frequency domain.

\begin{figure}[h]
\includegraphics[angle=0.0, width=8cm]{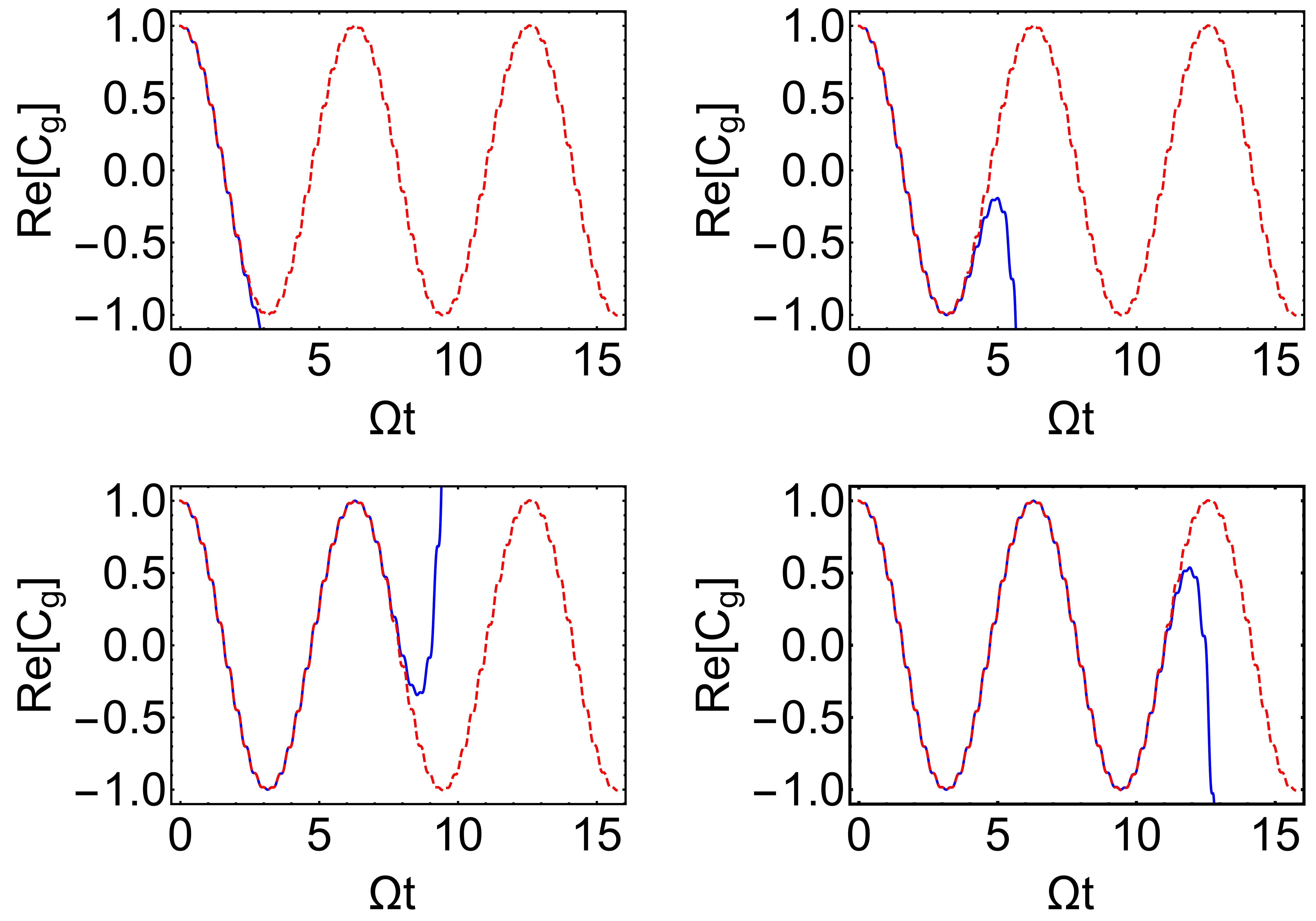}
\caption{(Color online) 
Comparison (for ${\rm Re}[C_g]$) between the numerical solution of
the semiclassical Rabi model (dashed red line) 
and the Picard series (solid blue line) up to third
(top left), eleventh (top right), twenty-first (bottom left), and
thirty-third (bottom right) order, for $\Omega/\omega=0.1$.}
\label{fig:noRWAsemiclassical}
\end{figure}

An example of the comparison between the exact (numerical)
solution of the semiclassical Rabi model and the truncated Picard
series is shown in Fig.~\ref{fig:noRWAsemiclassical}.
It is clear that the Picard expansion is suitable only for short 
times. Indeed, with expansion up to thirty-third order we can
faithfully reproduce the exact dynamics only up to less than two 
Rabi periods. From these plots we can also appreciate small 
(beyond RWA) oscillations, superposed to the main Rabi oscillations. 
The amplitude and frequency of these small oscillations will be
discussed in Sec~\ref{sec:Maghnussemi}.

\subsection{Quantum Rabi model}

\label{sec:PicardQRM}

\begin{figure}[h]
\includegraphics[angle=0.0, width=7cm]{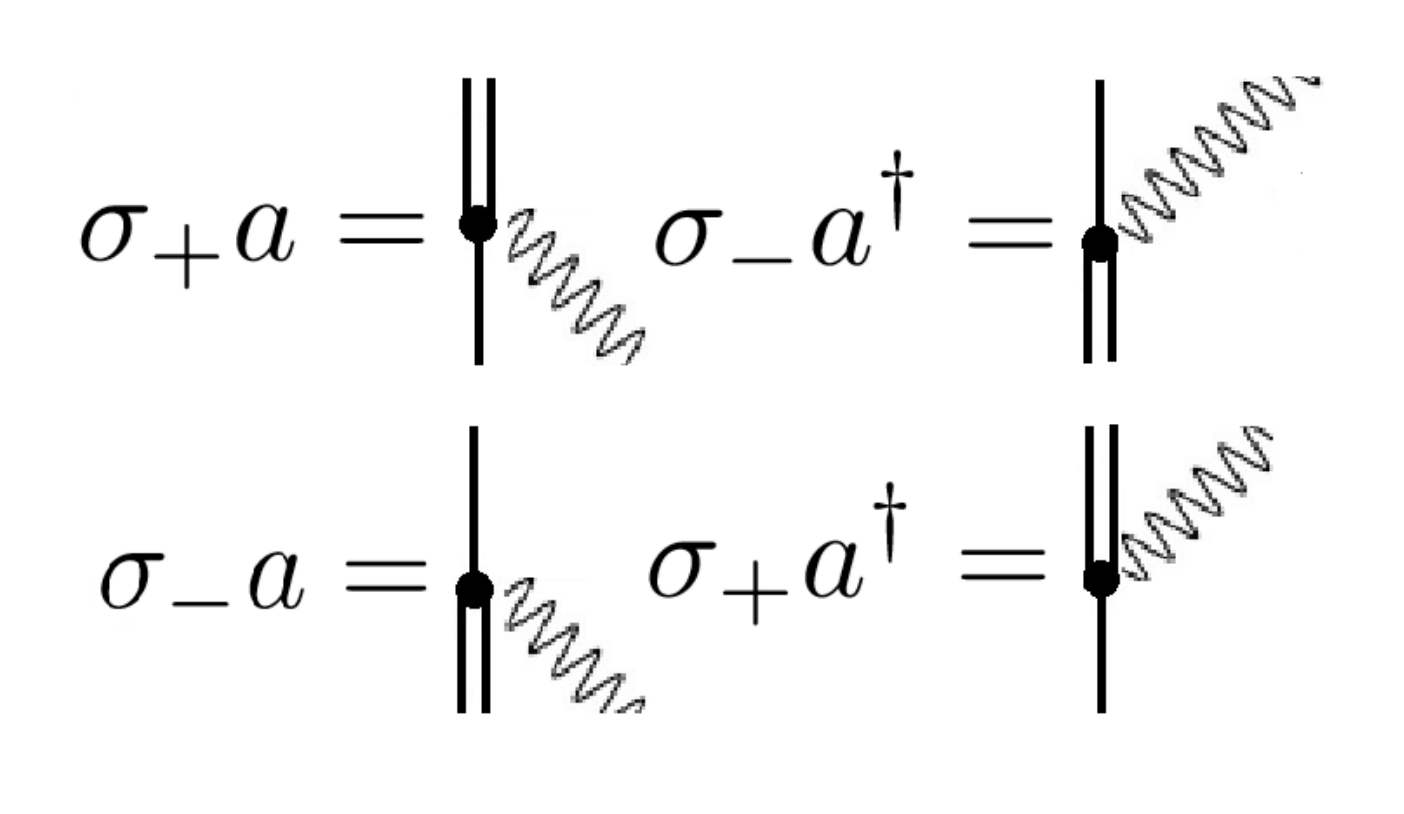}
\caption{
Vertices associated to the terms in Hamiltonian
(\ref{eq:HIquantum}). The vertices in the bottom line 
correspond to terms
neglected within the RWA.}
\label{fig:vertices}
\end{figure}

In this subsection, we review with more details the Picard expansion
for the finite-time quantum Rabi model introduced in Ref.~\cite{exotic}.
We expand the state vector in the $\{|l,n\rangle\}$ basis 
($l=g,e$; $n=0,1,2,...$) as 
$|\psi(t)\rangle=\sum_{l,n} C_{l,n}(t) |l,n\rangle$.
For every term in the Hamiltonian (\ref{eq:HIquantum}) it is possible
to give a diagrammatic representation (see Fig.~\ref{fig:vertices}).
The interaction vertex is represented by a full circle, a photon
by a wavy line, the qubit in the ground (excited) state by a straight
line (two parallel straight lines). Time flows from bottom to top. 
The vertex corresponding to the term proportional 
to $\sigma_+ a$ in the Hamiltonian tells us that we start from the
qubit in the ground state and a photon. As a consequence of the 
qubit-field interaction, the photon is absorbed
and the qubit is promoted to its excited state. 
The term $\sigma_- a^\dagger$ de-excites the atom while creating 
a photon, 
$\sigma_- a$ simultaneously destroys a photon 
and de-excites the atom,
and $\sigma_+ a^\dagger$ simultaneously creates a photon and 
excites the atom. 
The last two terms are responsible of the
anti-DCE and DCE effect, respectively.

We focus on the initial condition $|\psi(0)\rangle=|g,0\rangle$,
corrsponding to both the qubit and the field in their ground state.
Within the RWA, which conserves the total number of excitations
$N_T = \sigma_+ \sigma_- + a^\dagger a$, no excitations are possible
and $|\psi(t)\rangle=|g,0\rangle$ at all times.
On the other hand, the dynamics is nontrivial when the terms beyond 
RWA are included, since one can simultaneously excite the qubit
and create a photon, $\sigma_+ a^\dagger |g,0\rangle=|e,1\rangle$. 
The generation of photons from the vacuum is due to the 
nonadiabatic change of a system parameter (switching
of the qubit-field coupling constant) and is a manifestation
of the (parametric) DCE \cite{dodonov}.

\begin{figure}[h]
\includegraphics[angle=0.0, width=8cm]{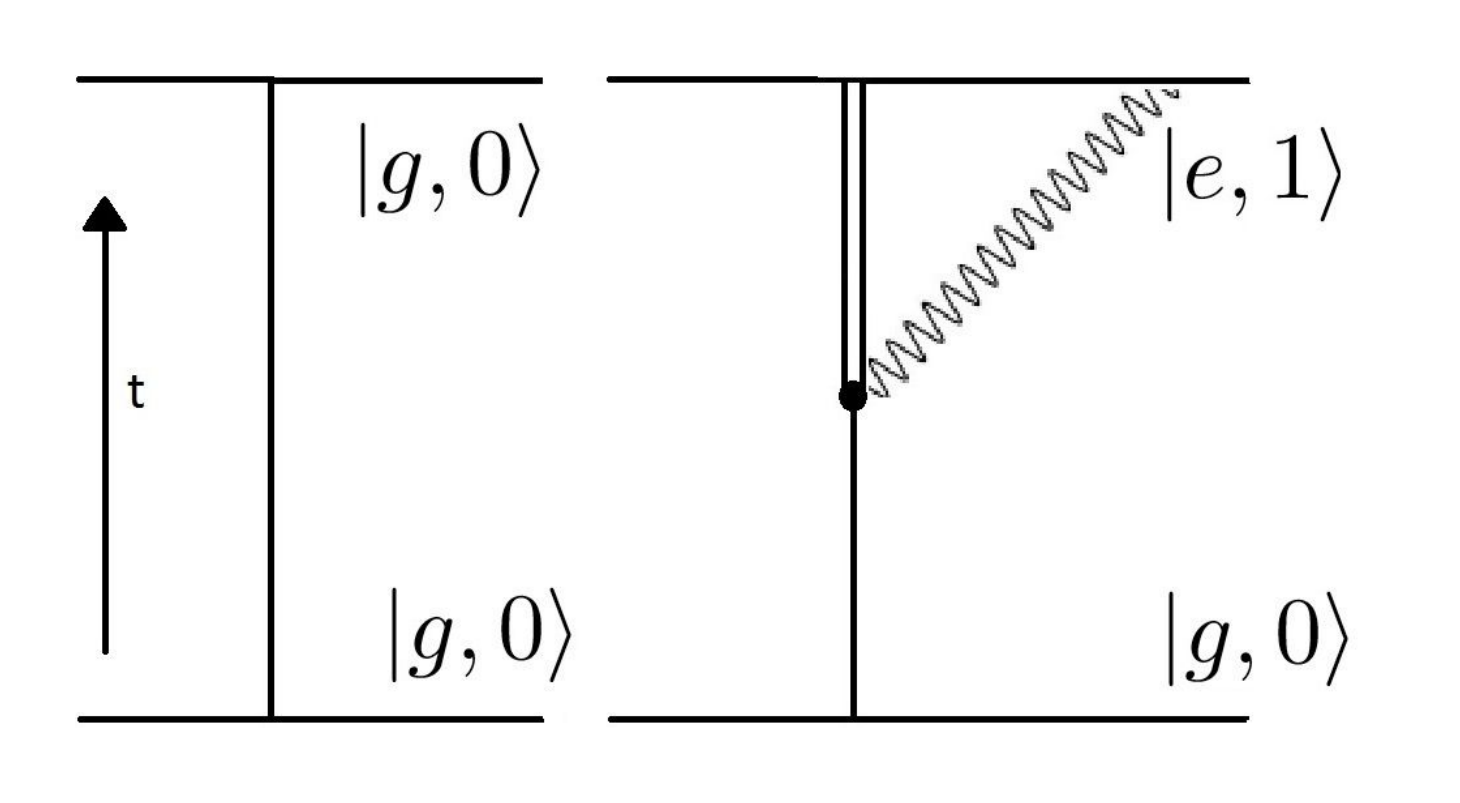}
\caption{
Diagrammatic representation of the zeroth- (left) and first-order 
(right) contributions in the Picard series for the quantum Rabi
model, with initial condition $|\psi(0)\rangle=|g,0\rangle$.}
\label{fig:order0-1}
\end{figure}

To the zeroth-order approximation
$|\psi^{(0)}(t)\rangle=|g,0\rangle$.
Such state is diagrammatically represented as a vertical
single line (see the left diagram in Fig.~\ref{fig:order0-1}),
meaning that the qubit remains in its ground state
$|g\rangle$, while no photons are emitted.
The two horizontal lines in Fig.~\ref{fig:order0-1} (left)
(as well as in all other diagrams in this paper)
mean that interaction is switched on at time $t=0$
(lower line) and switched off at time $t=\tau$ (upper line)
That is, these lines outline the fact that we are
dealing with finite-time QED.

To compute the first-order terms, we first observe that
$H_I(t') |\psi^{(0)}(t')\rangle=  e^{2i\omega t'}
\sigma_+ a^\dagger |g,0\rangle =e^{2i\omega t'}|e,1\rangle$.
After integrating 
$H_I(t') |\psi^{(0)}(t')\rangle$
from $t'=0$ to $t'=t$ 
according to Eq.~(\ref{eq:firstorder}),
we obtain
\begin{equation}
|\psi^{(1)}(t)\rangle=
\frac{\lambda}{2\omega}\,\left(1-e^{2i\omega t}\right)
|e,1\rangle.
\end{equation}
The diagrammatic representation of the first-order
contribution is shown in Fig.~\ref{fig:order0-1} (right):
the system starts from the state $|g,0\rangle$ and
performs a transition to the state $|e,1\rangle$, with
the qubit left in the excited state $|e\rangle$ and 
the emission of a single (real) photon. 
Note that this diagram is beyond the RWA, since the energy
is not conserved: both the qubit and the field start from
their ground states and are eventually excited.

\begin{figure}[h]
\includegraphics[angle=0.0, width=8cm]{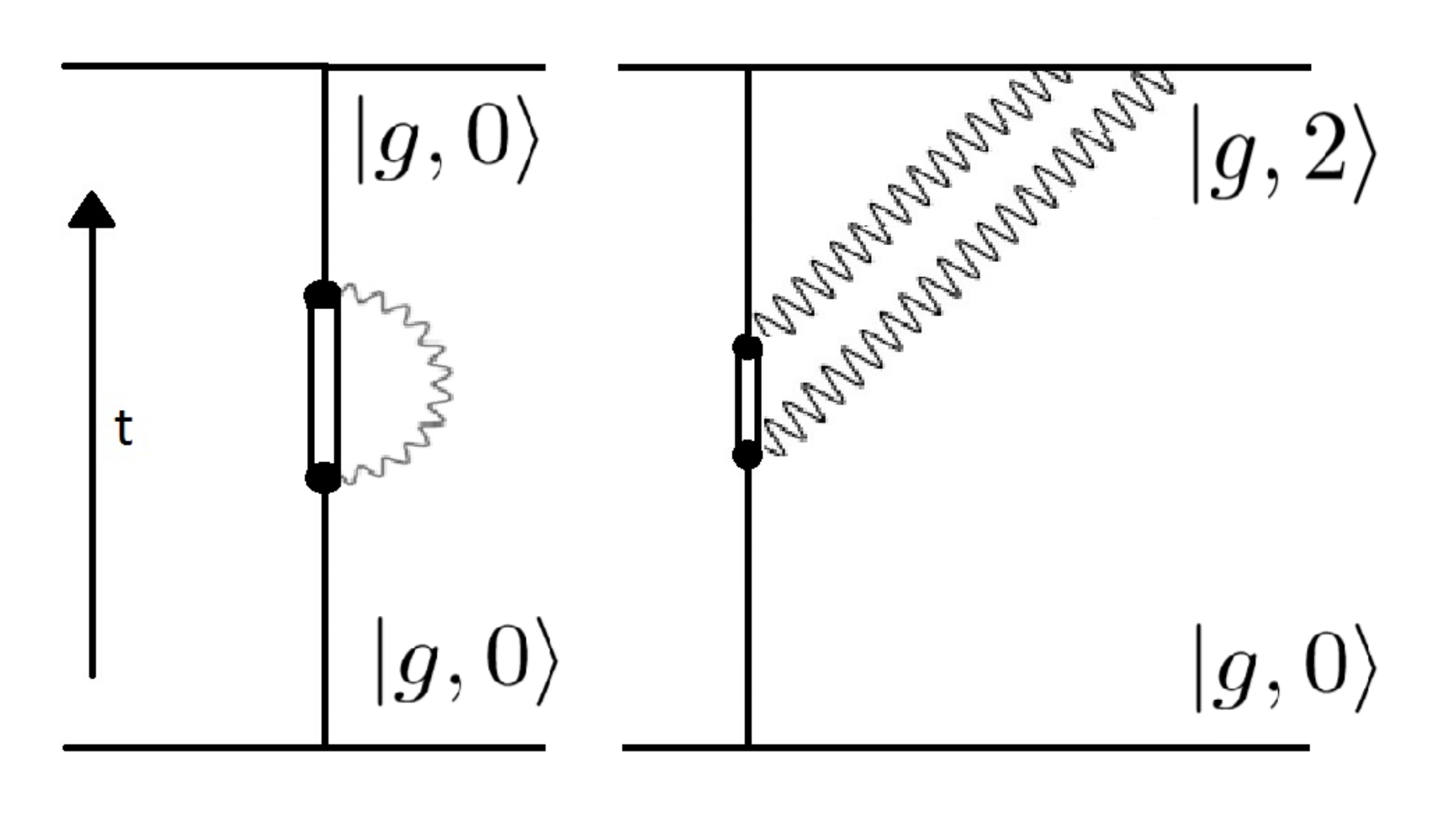}
\caption{
Same as in Fig.~\ref{fig:order0-1}, but for the second-order
contribtions.}
\label{fig:order2}
\end{figure}

To obtain the second-order contributions, we apply 
$H_I(t')$ to the first-order correction 
$|\psi^{(1)}(t')\rangle$. Since $|\psi^{(1)}(t')\rangle\propto 
|e,1\rangle$, we obtain terms proportional to 
$\sigma_- a |e,1\rangle= |g,0\rangle$ and 
$\sigma_- a^\dagger |e,1\rangle= \sqrt{2}\,|g,2\rangle$.
These contributions are represented by the diagrams of 
Fig.~\ref{fig:order2}.
Note that in the first case (left diagram)
the photon is virtual, while in the second
(right diagram) two real photons are emitted.
After integrating over time
$H_I(t') |\psi^{(1)}(t')\rangle$ according to 
Eq.~(\ref{eq:nthorder}) (with $n=2$),
we obtain
\begin{eqnarray}
\begin{array}{c}
{\displaystyle
|\psi^{(2)}(t)\rangle =
i\,\frac{\lambda^2}{2\omega}\left[
t+\frac{i}{2\omega}\left(1-e^{-2i\omega t}\right)\right]|g,0\rangle}
\\
{\displaystyle +\,
i\,\frac{\sqrt{2} \lambda^2}{2\omega}\left[
-t+\frac{i}{2\omega}\left(1-e^{2i\omega t}\right)\right]|g,2\rangle.}
\end{array}
\label{eq:dia2}
\end{eqnarray}
It is interesting to remark that in the latter term the
$\sqrt{2}$ factor is due to the stimulated emission
of the second photon by the first one.

\begin{figure}[h]
\includegraphics[angle=0.0, width=8cm]{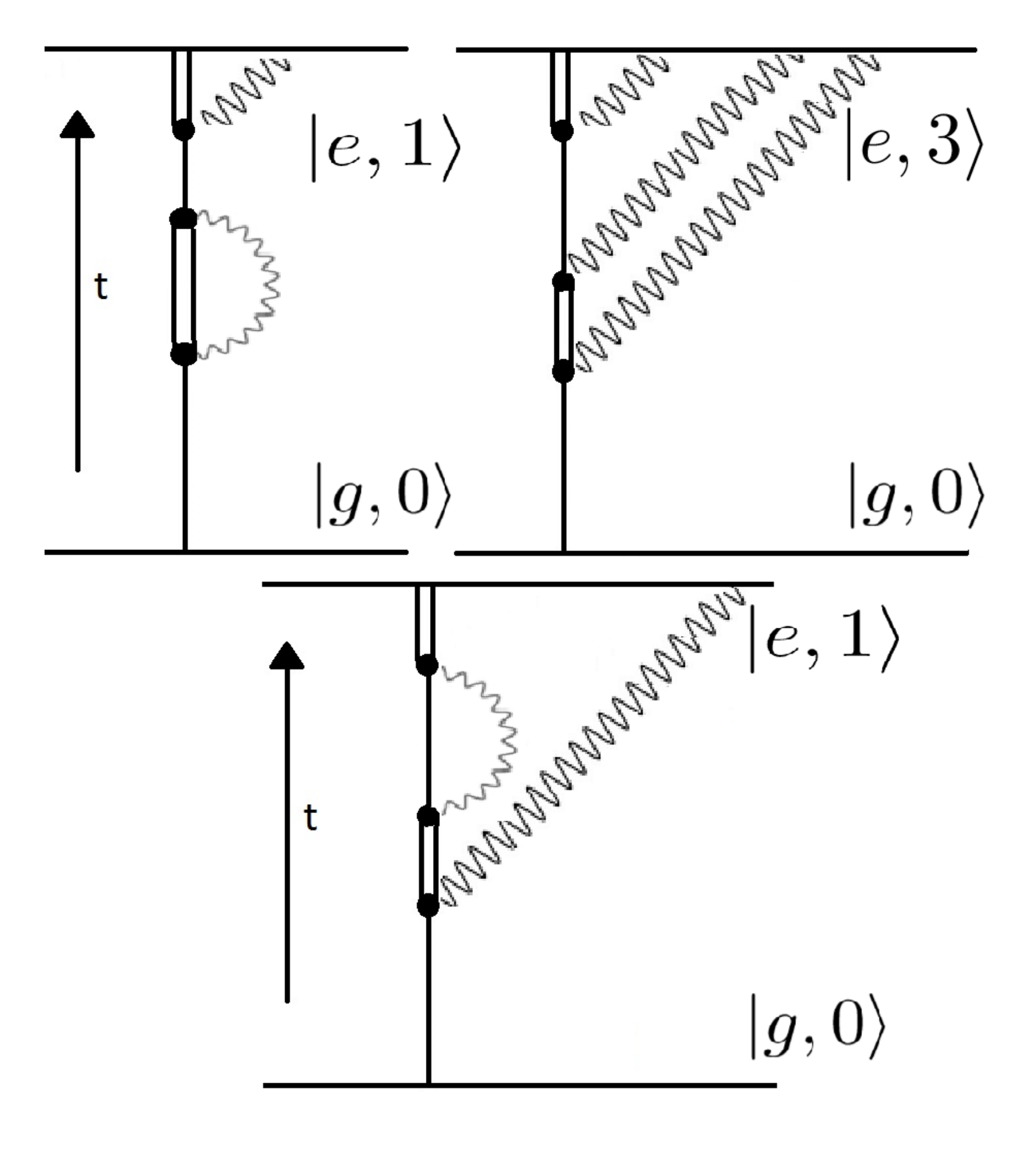}
\caption{
Same as in Fig.~\ref{fig:order0-1}, but for the third-order
contribtions.}
\label{fig:order3}
\end{figure}

To obtain the third-order contribution, we apply $H_I(t')$ 
to $|\psi^{(2)}(t')\rangle$. As a result, from the term
proportional to $|g,0\rangle$ in $|\psi^{(2)}(t')\rangle$
we obtain a term proportional to $|e,1\rangle$ (top left diagram
in Fig.~\ref{fig:order3}), while from the term 
proportional to $|g,2\rangle$ we obtain two terms, 
one proportional to $|e,3\rangle$ (top right diagram
in Fig.~\ref{fig:order3}) and one to 
$|e,1\rangle$ (bottom diagram
in Fig.~\ref{fig:order3}). 
After integrating over time
$H_I(t') |\psi^{(2)}(t')\rangle$, we obtain
\begin{eqnarray}
\begin{array}{c}
{\displaystyle
|\psi^{(3)}(t)\rangle =
\frac{\lambda^{3}}{4  \omega^{3}} \left[ -  1 + e^{2 i \omega t}
- i  (\omega t) \left(1 + e^{2 i \omega t}\right)\right] |e,1\rangle 
}\\
{\displaystyle
+\sqrt{\frac{3}{2}}  \frac{\lambda^{3}}{8  \omega^{3}} \left[
1 - e^{4 i \omega t} + 4 i e^{2 i \omega t} (\omega  t)\right]|e,3\rangle 
}\\
{\displaystyle
+\frac{\lambda^{3}}{4  \omega^{3}} \left[1 - e^{2 i \omega t} +  
2 i (\omega\,t) -   2  (\omega t)^{2}\right] |e,1\rangle,
}
\end{array}
\end{eqnarray}
where the three terms of this equation correspond,
respectively, to the
top left, top right, and bottom diagram of Fig.~\ref{fig:order3}).
The perturbative treatment outlined in this subsection
can be easily iterated to higher orders.

\begin{figure}[h]
\includegraphics[angle=0.0, width=8cm]{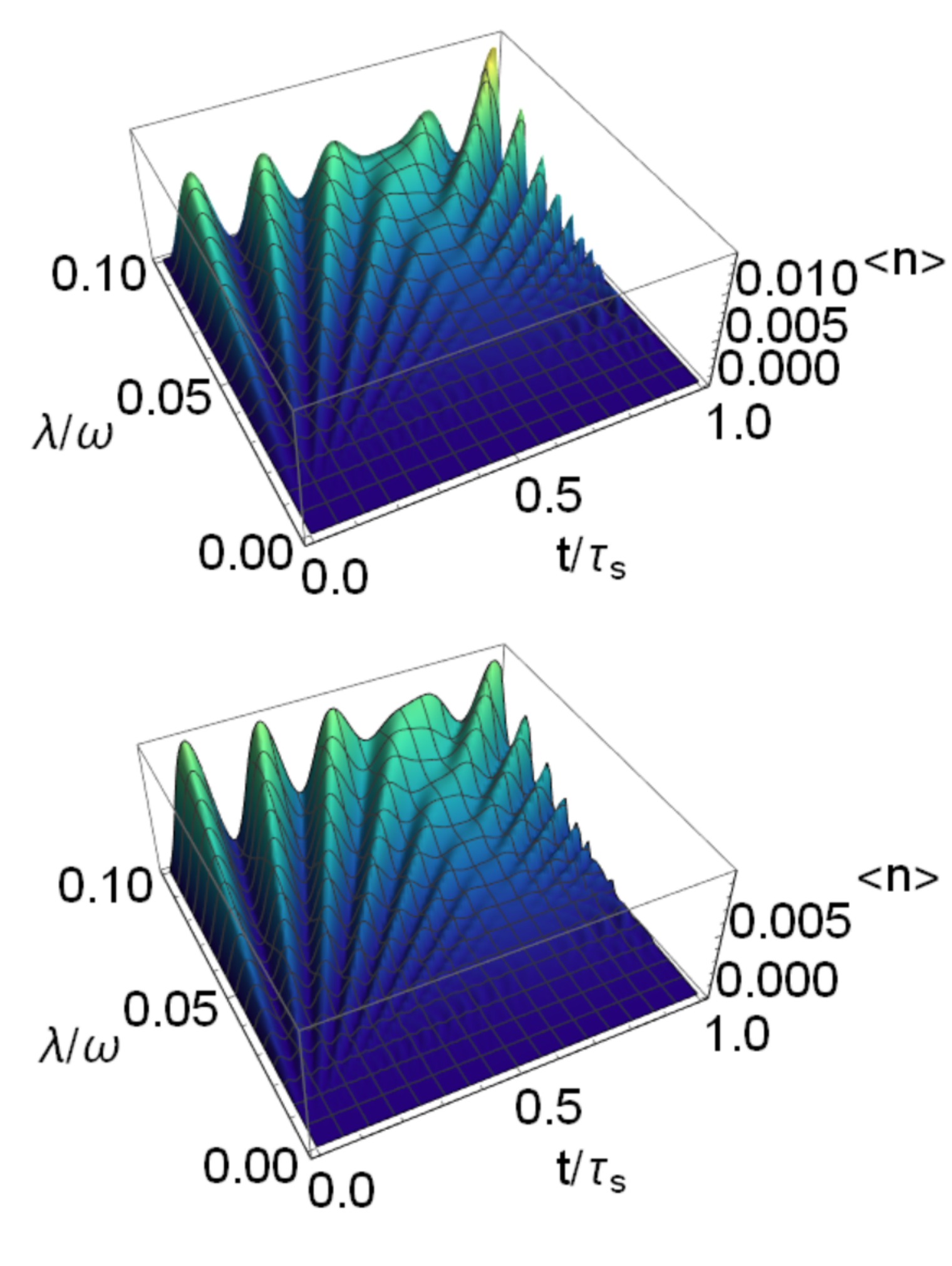}
\caption{(Color online) 
Mean number of generated photons $\langle n \rangle$ as a function of 
the coupling strength $\lambda$ (in units of $\omega$) 
and of the interaction time $t$,
measured in units of the swapping time $\tau_s = \pi/2 \lambda$.
The fourth-order Picard expansion (top) is compared with the 
numerical results (bottom).}
\label{fig:picard-3d}
\end{figure}

As an example of application of the Picard expansion,
we compute the mean number of generated photons, as a 
function of the qubit-field coupling constant $\lambda$ and of the
interaction time $t$. We can see from Fig.~\ref{fig:picard-3d} 
that the fourth-order plot (top panel) is in good agreement
with the exact solution (bottom panel), provided 
$\lambda$ and $t$ are not too large.

\begin{figure}[h]
\includegraphics[angle=0.0, width=8cm]{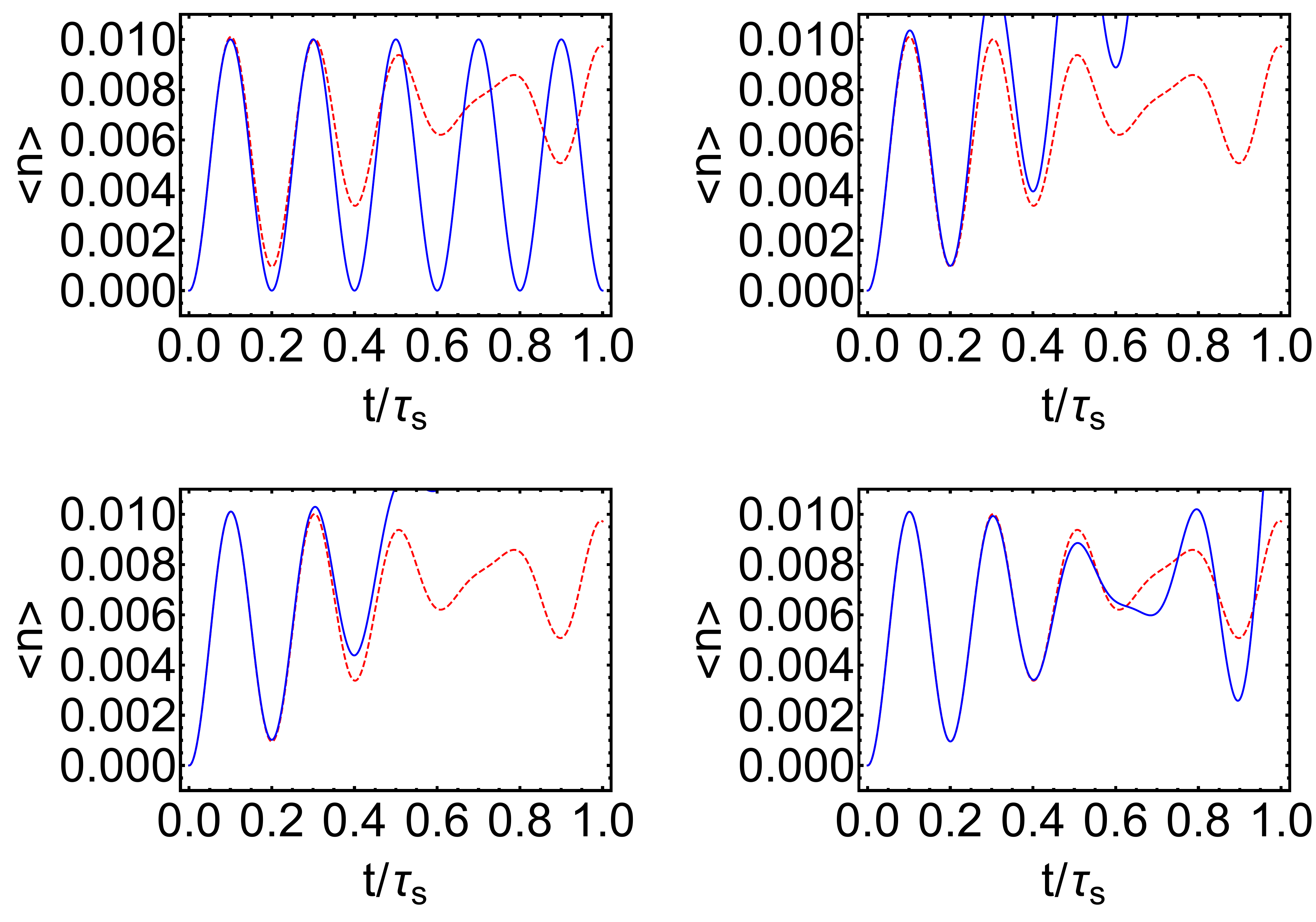}
\caption{(Color online) 
Mean number of photons as a function of time, with the numerical 
results (dashed red line) compared with the Picard 
expansion (solid blue line) truncated to the first 
(top left), second (top right), third (bottom left), and
fourth (bottom right) order, for $\lambda/\omega=0.1$.}
\label{fig:npicardexact}
\end{figure}

\begin{figure}[h]
\includegraphics[angle=0.0, width=7.5cm]{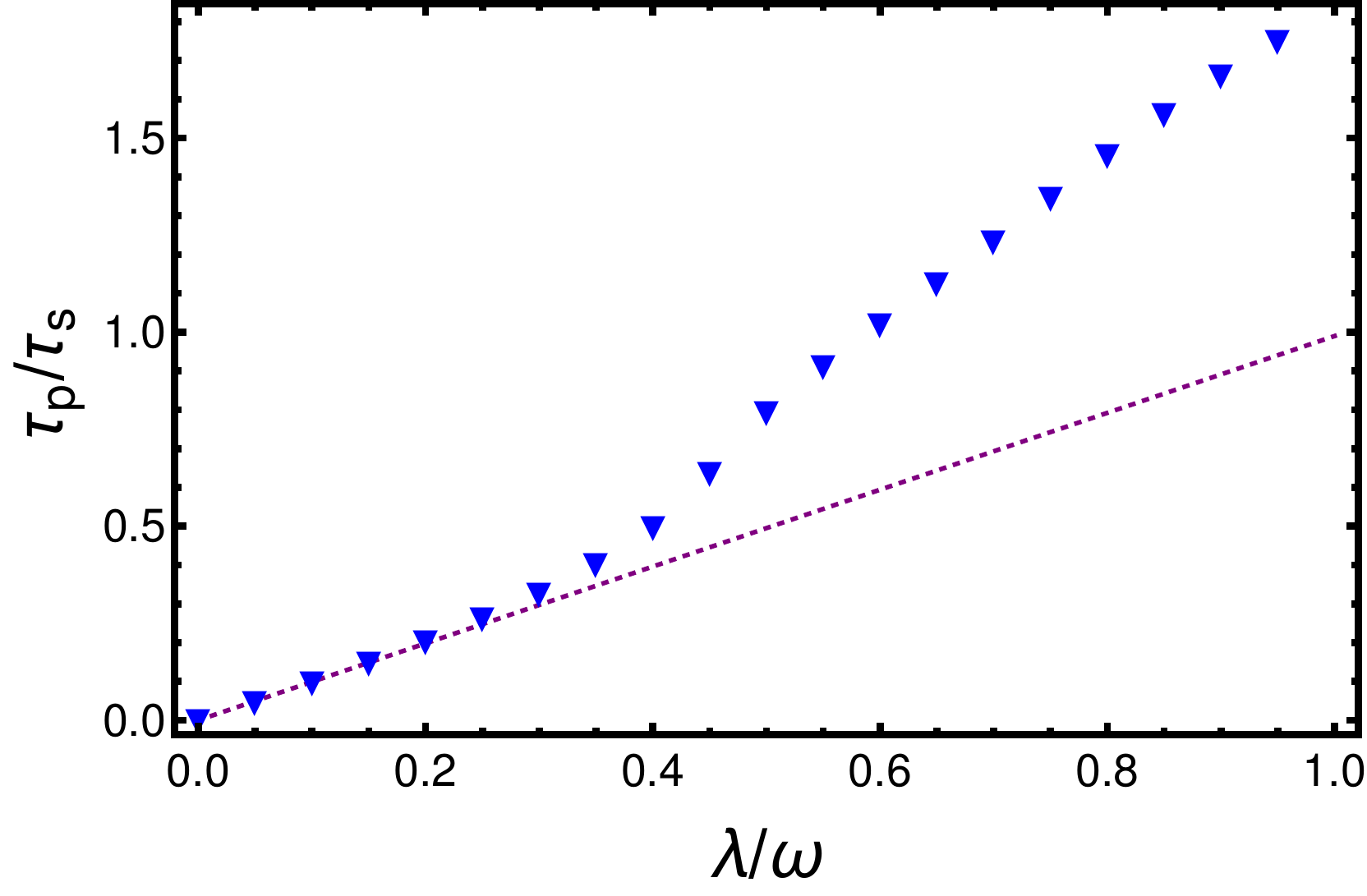}
\caption{(Color online) 
Time $\tau_p$ of the first peak in $\langle n \rangle (t)$:
comparison of the first-order (dashed line) with 
the numerical results (triangles).}
\label{fig:npertexact}
\end{figure}

For $\lambda/\omega=0.1$,
the Picard expansion is compared (up to the fourth order) 
with the exact numerical solution in Fig.~\ref{fig:npicardexact}.
It can be seen that the Picard series
truncated to the fourth order can reproduce the behavior of 
$\langle n \rangle$ up to $t/\tau_s\approx 0.5$. 
On the other hand, for this value of $\lambda$ the amplitude
and time of the first peak can be estimated already from the 
first-order expansion. To the first order,
\begin{equation}
\langle n \rangle (t) =\left(\frac{\lambda}{\omega}\right)^2
\sin^2(\omega t),
\label{eq:taup}
\end{equation}
corresponding to the first peak at time $\tau_p$, with 
$\tau_p/\tau_s=\lambda/\omega$, and peak value 
$\langle n\rangle (\tau_p)=(\lambda/\omega)^2$.
As shown in Fig.~\ref{fig:npertexact}, this analytical
prediction for $\tau_p$ is in good agreement with the 
numerical results up to $\lambda/\omega\approx 0.3$. 

The oscillations in $\langle n \rangle (t)$, due to the 
coherent generation (DCE) and destruction (anti-DCE) of 
photons, are a clear dynamical ``smoking gun'' of the 
ultrastrong coupling regime. Such oscillations, as shown
in Figs.~\ref{fig:picard-3d} and \ref{fig:npicardexact}, are, 
for relatively small
values of $\lambda/\omega$, regular. At small times, 
a quasi-periodic behavior with frequency $2\omega$ 
is clearly seen, and also predicted by first-order 
perturbation theory, Eq.~(\ref{eq:taup}).
This reasult might be interesting for experimental investigations 
in that clear features of the DCE
are observable with short interaction times and relatively
small interaction strengths.

\section{Magnus expansion}

The Magnus expansion starts by assuming that an exponential 
form for the (unitary) time-evolution operator $U(t)$ 
(defined by $|\psi(t)\rangle=U(t)|\psi(0)\rangle$) exists:
\begin{equation}
U(t)=e^{\Omega(t)}, \quad \Omega(0)=0,
\label{eq:magnus1}
\end{equation}
with a series expansion for $\Omega$:
\begin{equation}
\Omega(t)=\sum_{n=1}^\infty \Omega^{(n)}(t).
\label{eq:magnus2}
\end{equation}
An approximate expression for the time-evolution operator is
obtained by truncation of the Magnus expansion. 
The first few terms in such expansion 
are 
\begin{eqnarray}
\begin{array}{c}
{\displaystyle
\Omega^{(1)}(t)=\int_0^t dt_1 A(t_1),
}
\\
{\displaystyle
\Omega^{(2)}(t)=\frac{1}{2}\int_0^t dt_1 
\int_0^{t_1} dt_2 [A(t_1),A(t_2)],
}
\\
{\displaystyle
\Omega^{(3)}(t)=\frac{1}{6}\int_0^t dt_1
\int_0^{t_1} dt_2 \int_0^{t_2} dt_3
}
\\
{\displaystyle
([A(t_1),[A(t_2),A(t_3)]]+[A(t_3),[A(t_2),A(t_1)]]),
}
\end{array}
\label{eq:Magnusgeneric}
\end{eqnarray}
where we have defined the (anti-Hermitian) operator 
$A(t)=-iH_I(t)$, with $H_I$ Hamiltomian in the interaction picture. 
For a derivation of the terms $\Omega^{(n)}(t)$ see, 
\emph{e.g.}, Ref.~\cite{Blanes2009}. 
Note that, since the expansion is for $\Omega$
and not for $U$ as in the Picard series, 
the Magnus expansion provides a \emph{unitary perturbation theory}, 
in contrast to the Picard series.
This is one of the most appealing features of the Magnus expansion.
The Magnus expansion, 
in particular conditions for the convergence of the 
Magnus series and several applications of the method, 
including its use as a numerical integrator, are 
reviewed in Ref.~\cite{Blanes2009}.  
Hereafter, we shall discuss applications of the Magnus expansion
to the Rabi model. 

\subsection{Semiclassical Rabi model}
\label{sec:Maghnussemi}

We write explicitly the first three terms of the Magnus expansion 
for the semiclassical Rabi model. 
Let $\Omega^{(n)}_{ij}=\langle i | \Omega^{(n)}|j\rangle$,
with $i,j=g,e$, denote the matrix elements of $\Omega^{(n)}$ 
in the $\{|g\rangle,|e\rangle\}$ basis. 
From Eq.~(\ref{eq:Magnusgeneric}), using the semiclassical 
Rabi Hamiltonian (\ref{eq:HIsemi}) we obtain
\begin{eqnarray}
\begin{array}{c}
{\displaystyle
\Omega^{(1)}_{gg}(t)=\Omega^{(1)}_{ee}(t)=0,
}
\\
{\displaystyle
\Omega^{(1)}_{ge}(t)= -\frac{\Omega}{2\omega} 
        \left( 1-e^{-2i\omega t} +2i\omega t \right)
=-[\Omega^{(1)}_{eg}(t)]^\star,
}
\end{array}
\end{eqnarray}
\begin{eqnarray}
\begin{array}{c}
{\displaystyle
\Omega^{(2)}_{ge}(t)=\Omega^{(2)}_{eg}(t)=0,
}
\\
{\displaystyle
\Omega^{(2)}_{gg}(t)= 
\frac{i \Omega^2}{4 \omega^2} 
\left( - 2\omega t \cos(2\omega t)+
\sin(2\omega t)   \right)
}
\\
{\displaystyle
=-[\Omega^{(2)}_{ee}(t)],
}
\end{array}
\end{eqnarray}
\begin{eqnarray}
\begin{array}{c}
{\displaystyle
\Omega^{(3)}_{gg}(t)=\Omega^{(3)}_{ee}(t)=0,
}
\\
{\displaystyle
\Omega^{(3)}_{ge}(t)= 
 \frac{\Omega^3}{8 \omega^3}  
\left[
- 3  + i  \omega  t + \frac{4}{3} \, \omega^{2}  t^{2} \right.
}
\\
{\displaystyle
+\left( \frac{3}{2} + 2  i  \omega  t - \frac{2}{3} 
\,\omega^{2}  t^{2}  \right)  e^{-2i\omega t} 
}
\\
{\displaystyle
+ \left( \frac{7}{6} - \frac{4}{3} \, i  \omega  t 
- \frac{2}{3} \,\omega^{2}  t^{2}  \right)  e^{2i\omega t} 
}
\\
{\displaystyle
\left.
+\left( \frac{1}{3} + \frac{1}{3} \, i  \omega  t  \right) 
\, e^{-4i\omega t}   \right]
=-[\Omega^{(3)}_{eg}(t)]^\star.
}
\end{array}
\end{eqnarray}

Within the RWA, the semiclassical Rabi model is, in the interaction 
picture, time-independent, and therefore the Magnus expansion reduces
to its first-order term, $\Omega(t)=\Omega^{(1)}(t)=-i H_I t$. 
On the other hand, when the terms beyond RWA are taken into account, 
in general $[A(t_1),A(t_2)]\ne 0$ if $t_1\ne t_2$ and therefore
we must consider also higher-order terms in the Magnus expansion. 

\begin{figure}[h]
\includegraphics[angle=0.0, width=8.5cm]{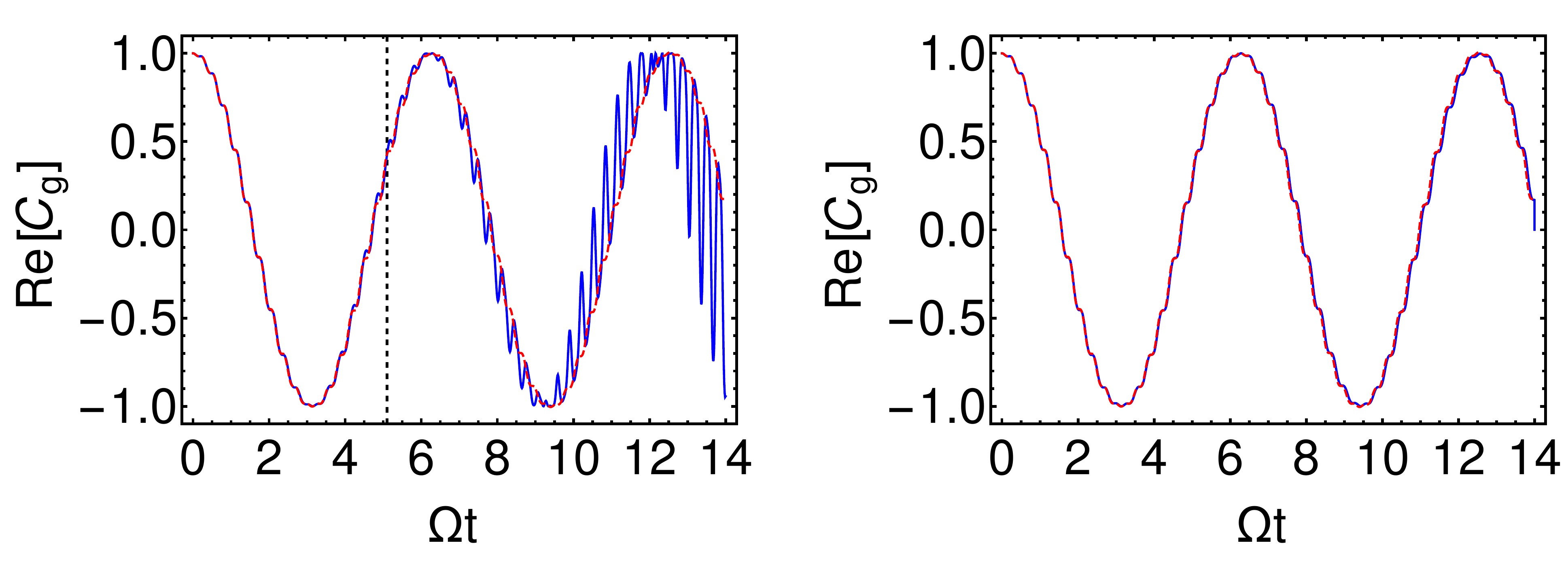}
\caption{(Color online) 
Comparison (for ${\rm Re}[C_g]$) between the numerical solution of
the semiclassical Rabi model (dashed red line) 
and the Magnus expansion (solid blue line), up to fourth order
(left) or  iterating five times the first order 
expansion (right), for $\Omega/\omega=0.1$. The dashed line shows the time
($\Omega t\approx 5.1$)
up to which convergence of the Magnus expansion is guaranteed by 
criterion (\ref{eq:magnusconvergence}).}
\label{fig:magnusrk}
\end{figure}

As an example, in Fig.~\ref{fig:magnusrk} (left panel) we compare 
the Magnus expansion, truncated to the fourth order, with the
numerical integration of the Schr\"odinger equation via a
fourth-order Runge-Kutta method. If we compare these results
with those obtained by means of the Picard series
(see Fig.~\ref{fig:noRWAsemiclassical}), it is clear that the
Magnus expansion allows us to address much longer evolution times already 
at small orders. 

On the other hand, the convergence of the
Magnus expansion is not guaranteed at all times. More precisely,
a sufficiently condition \cite{moan,casas2007} 
for the convergence of the Magnus 
expansion is that
\begin{equation}
\int_0^t ||A(t')||_2 d t' < \pi,
\label{eq:magnusconvergence}
\end{equation}
where $||A||_2$ is the square root of the largest eigenvalue of
$A^\dagger A$.
In the example of Fig.~\ref{fig:magnusrk}, this criterion 
ensures convergence for times $t<t_c$, with $\Omega t_c\approx 5.1$
(vertical dashed line in the figure). 
For $t>t_c$, the strong oscillations and the discrepancy 
between the Magnus expansion truncated to
the fourth-order and the exact numerical solution, suggest a 
different numerical approach. That is, we concatenate truncated 
Magnus expansions. With this approach, we can address arbitrarily
long time scales. For instance, Fig.~\ref{fig:magnusrk} (right panel)
shows the good agreeement between the numerical solution and 
the concatenation of five first-order Magnus expansions. 

\begin{figure}[h]
\includegraphics[angle=0.0, width=8.5cm]{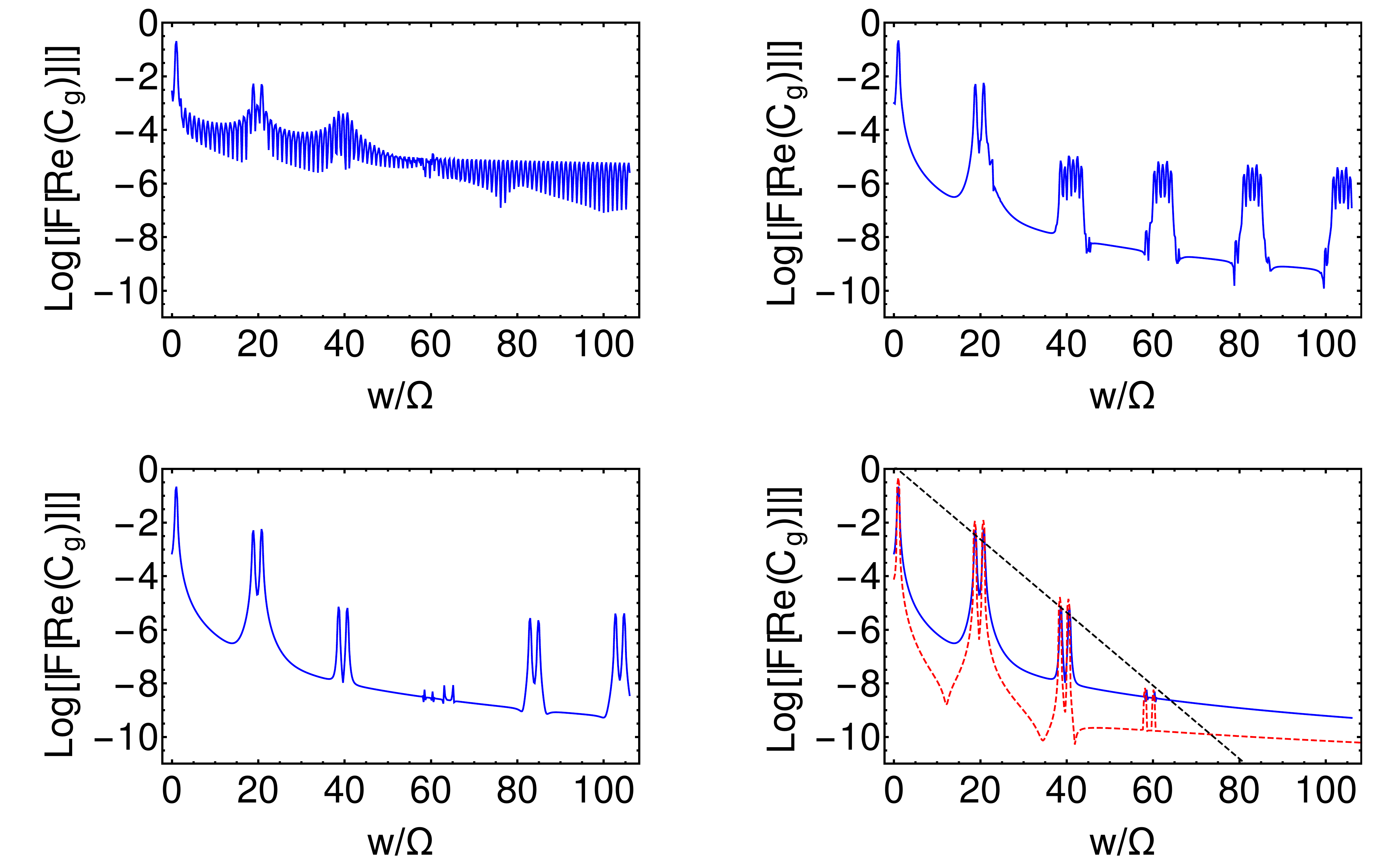}
\caption{(Color online) 
Fourier transform $F$ of ${\rm Re}({C}_g)$
(arbitrary units in the plot), 
obtained from integration of the semiclassical 
Rabi model up to $\Omega t=60 \pi$, with 
$\Omega/\omega=0.1$, iterating the 
fourth-order Magnus expansion ${\cal N}=5$ (top left), 100 (top right),
500 (bottom left), and 3000 (bottom right) times.
The dashed red curve in the bottom right panel is instead obtained
by fourth-order Runge-Kutta integration of the equations of 
motion, with $5.12\times 10^5$ points. 
The dashed line ${\rm Log}[|F[{\rm Re}(C_g)]|]=a-b(w/\Omega)$, with 
$a\approx 0.106$ and $b\approx 0.136$, fits the decay of the
peaks in the Fourier transform.} 
\label{fig:fft-semiclassical}
\end{figure}

To further assess the validity of the Magnus expansion, we 
follow the dynamics up to 30 Rabi periods 
($\Omega t=60\pi$), by concatenating ${\cal N}$ times the 
fourth-order Magnus expansion, and then compute the Fourier 
transforms $F$ of $C_g$ and $C_e$. 
As an example, we show in 
Fig.~\ref{fig:fft-semiclassical} 
$F[{\rm Re}({C}_g)]$,
for different values of ${\cal N}$. We can see that 
${\cal N}=3\times 10^3$ allow us to reproduce the main 
features of the Fourier spectrum: for that purpose,  
more than $10^5$ time steps are necessary
when using the Runge-Kutta method
(see the bottom right panel of Fig.~\ref{fig:fft-semiclassical}).
The Magnus expansion can then be used as a numerical integrator,
more efficient for this problem than the Runge-Kutta mehod,
as it allows much longer time steps. 

The Fourier spectrum has characteristic double-peaks.
More precisely, Fig.~\ref{fig:fft-semiclassical} exhibits 
a single peak at the Rabi frequency $\Omega$,
and doublets at frequencies $2n\omega \pm  \Omega$, with 
$n=1,2,3,...$. 
Such features can be qualitatively explained as follows. 
The peak at frequency $\Omega$ corresponds to Rabi oscillations
and already exists within the RWA. On the other hand, the doublets
are structures beyond RWA, which can be conveniently understood 
from the Picard series. At each order of the Picard expansion,
we integrate in time terms proportional to 
$e^{\pm 2i\omega t}$ times the wave-function at the previous 
order. We therefore generate new harmonics at higher 
frequency as we increase the perturbative order in the Picard
series.
Terms proportional to $e^{\pm 2 i  n \omega t}$ multiply the 
Rabi oscillations, proportional to $e^{i\Omega t}$, 
and therefore in conclusion we generate 
harmonics at frequencies $2 n\omega \pm \Omega$.   
Note that each integration in time of $e^{\pm 2i\omega t}$ implies 
a decay of the weight of the corresponding harmonic by a factor 
$1/(2\omega)$. 
If we write the Schr\"odinger equation for the semiclassical Rabi model 
(\ref{eq:HIsemi}) as
\begin{equation}
\left [
\begin{array}{c}
{\dot C}_{g}(t) \\
{\dot C}_{e}(t)
\end{array}
\right ] = -  i f(t)
\left [
\begin{array}{cc}
0 & 1 + e^{- 2 i \omega t} \\
 1 + e^{ 2i\omega t}   & 0
\end{array}
\right ]
\left [
\begin{array}{c}
 C_{g}(t) \\
 C_{e}(t)
\end{array}
\right ],
\label{eq:schrodingerRabi}
\end{equation}
we can clearly see that at each order of the Picard series we 
improve the approximation for either $C_g$ or $C_e$.
Therefore, we need two steps in the Picard expansion to improve
$C_g$ (or $C_e$) and generate new harmonics. 
Since this implies two integrations in time, the harmonics 
at frequencies $2n\omega\pm \Omega$ are scaled by a 
factor $[\Omega/(2\omega)]^2$ with respect to the harmonics at frequencies
$2(n-1)\omega\pm \Omega$.
This estimate is in good agreement with the numerical results
of Fig.~\ref{fig:fft-semiclassical}.
Indeed, for $\Omega/\omega=0.1$ 
the decay of the first peaks in the Fourier transform
is fitted by an exponential law, 
${\rm Log}[|F[{\rm Re}(C_g)]|]=a-b(w/\Omega)$, with
$a\approx 0.106$ and $b\approx 0.136$. 
This implies that the ratio between the amplitude of
nearby doublets is approximately equal to 
$10^{-2(\omega/\Omega) b}\approx 1/525$, 
not far from $[\Omega/(2\omega)]^2=1/400$. 
A more precise calculation appears difficult, since
at each perturbative order new harmonics are generated
but also the weight of the already existing harmonics
is modified.
Note that 
in the Magnus series, since we have an exponential approximation 
theory (i.e., we consider $e^{\Omega}$, with a truncated 
expansion for $\Omega$), higher-order harmonics are visible 
already at the lowest orders.  

\begin{figure}[h]
\includegraphics[angle=0.0, width=8.5cm]{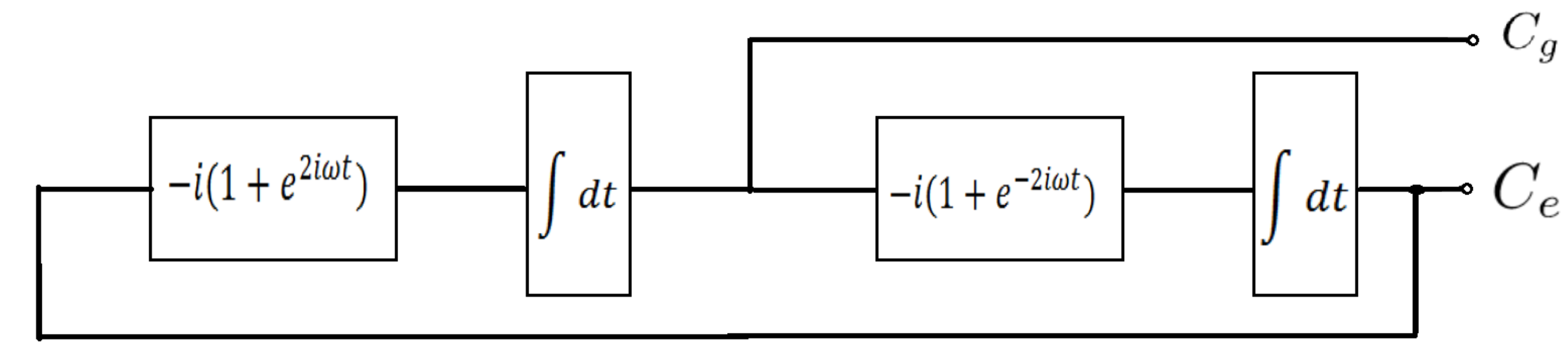}
\caption{Schematic drawing of an analog circuit for 
the integration of the Schr\"odinger equation
(\ref{eq:schrodingerRabi}) for the semiclassical Rabi model.}
\label{fig:circuitRabi}
\end{figure}

The above discussion can be visualized by means of the analog circuit 
reported in Fig.~\ref{fig:circuitRabi}. 
It corresponds to two orders
in the Picard expansion, and each integration brings a factor 
$\Omega/(2\omega)$. The signal ($C_g$ and $C_e$) can be reinjected
and at each loop the approximation is improved, adding each time two 
more orders in the Picard series. 

Finally, we point out that there is an interesting analogy,
in particular with respect to the occurrence of doublets, 
between the semiclassical Rabi model and the Mathieu 
equation in an appropriate range of parameters, 
see Appendix~\ref{app:mathieu}.

\subsection{Quantum Rabi model}

\begin{figure}[h]
\includegraphics[angle=0.0, width=8.5cm]{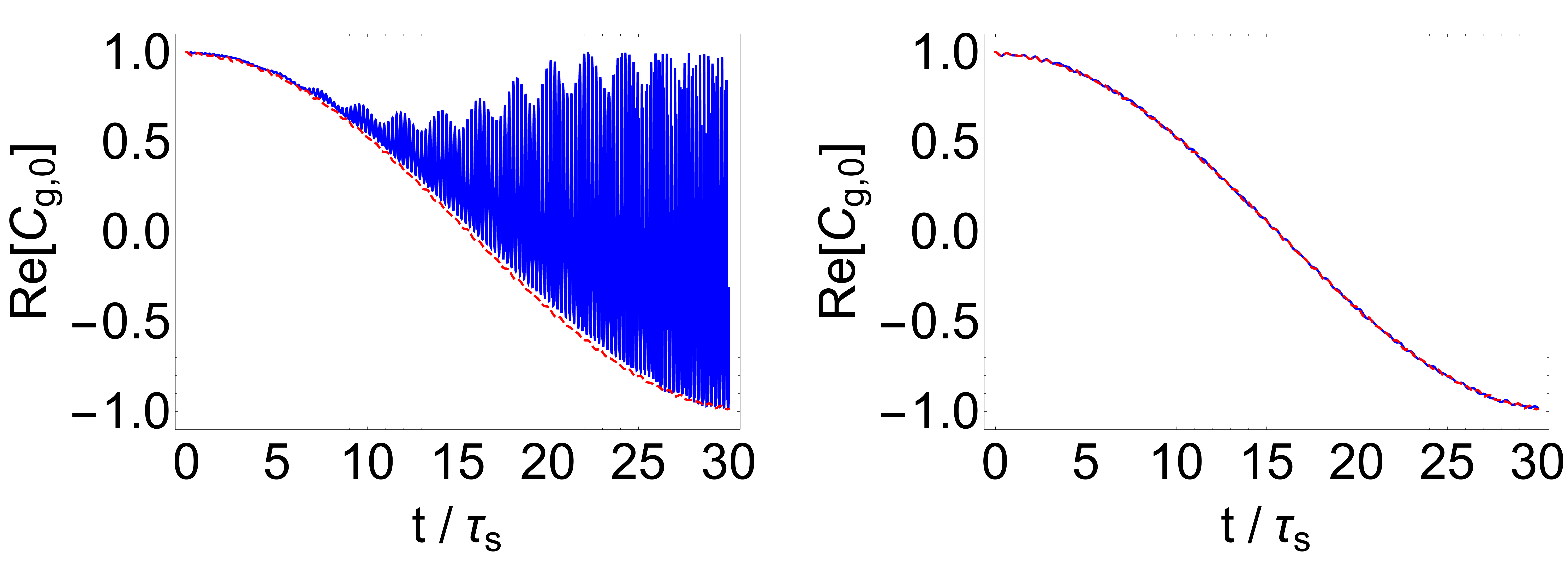}
\caption{(Color online) 
Comparison (for ${\rm Re}[C_{0g}]$) between the numerical solution of
the semiclassical Rabi model (dashed red line) 
and the Magnus expansion (solid blue line), up to fourth order
(left) or  iterating a hundred times the fourth order 
expansion (right), for $\lambda/\omega=0.12$.
Note that the initial condition we used, $C_{g,0}(t=0)=1$, is 
such that within the RWA the dynamics is trivial, 
$C_{g,0}(t)=1$ at all times.}
\label{fig:magnusqrk}
\end{figure}

The Magnus expansion can also be applied to the quantum 
Rabi model, using Eqs.~(\ref{eq:magnus1}), (\ref{eq:magnus2})
and (\ref{eq:Magnusgeneric}). For the sake of simplicity, we do 
not report explicit expressions for $\Omega^{(n)}(t)$.  
As the Hilbert space is infinite-dimensional, we cannot use convergence
criteria like Eq.~(\ref{eq:magnusconvergence}), since the eigenvalues
of $A^\dagger A$ are not upper bounded.
On the other hand, for any given initial condition the Hilbert space
actually explored by the dynamics is finite. 
For instance, if initially both the field and the qubit are prepared
in their ground state, as discussed in Sec.~\ref{sec:PicardQRM} 
the mean number of photons does not grow
indefinitely but oscillates 
due to coherent generation (DCE) and
destruction (anti-DCE) of photons. 
Hence, we expect convergence of the Magnus expansion
for sufficiently short integration times. Such expectation is borne
out by numerical data, as shown in Fig.~\ref{fig:magnusqrk}.

\section{Conclusions}

In this paper, we have applied the Picard and Magnus expansions 
to the ultrastrong matter-field coupling, in the paradigmatic 
Rabi model. The Picard series, truncated to low orders, 
is suitable only for short interaction times.
On the other hand, we have shown
that the Magnus expansion, through concatenation, is an efficient
numerical integrator, in that it allows time steps much longer than
in the Runge-Kutta method. 

We have highlighted clear features of the dynamics in the 
ultrastrong coupling regime, and in particular of the dynamical Casimir effect. 
Regular oscillations in the mean number $\langle n \rangle$
of photons, 
due to the coherent generation (DCE) 
and destruction (anti-DCE) of photons take place.
This reasult provides a clear ``smoking gun'' of the DCE,
which might be of interest for experimental investigations in circuit 
QED, in that the above oscillations are observable 
with short interaction times and relatively small interaction strengths.

We have shown that the Fourier spectrum of 
motion in the semiclassical Rabi model exhibits a peak at 
the Rabi frequency $\Omega$ and doublets
at frequencies $2n\omega\pm\Omega$, with $n$ positive integer.
While the Rabi frequency is trivially obtained by solving the 
Rabi model within the rotating wave approximation, 
the doublets are features beyond the RWA. 
Both the oscillations in $\langle n \rangle$ and the doublets 
can be explained by means of the Picard series.   
The Fourier analysis can be extended also to the quantum Rabi model
finding similar, even though more complicated structures with 
doublets. Finally, the analogy with the Mathieu equation highligths the fact 
that doublets are a general feature of time-modulated systems.

{\it Acknowledgments:} We acknowledge support by the INFN
through the project ``QUANTUM''. 

\appendix

\section{Analogy between the semiclassical Rabi model and the Mathieu 
equation}

\label{app:mathieu}

\begin{figure}[h]
\includegraphics[angle=0.0, width=8.5cm]{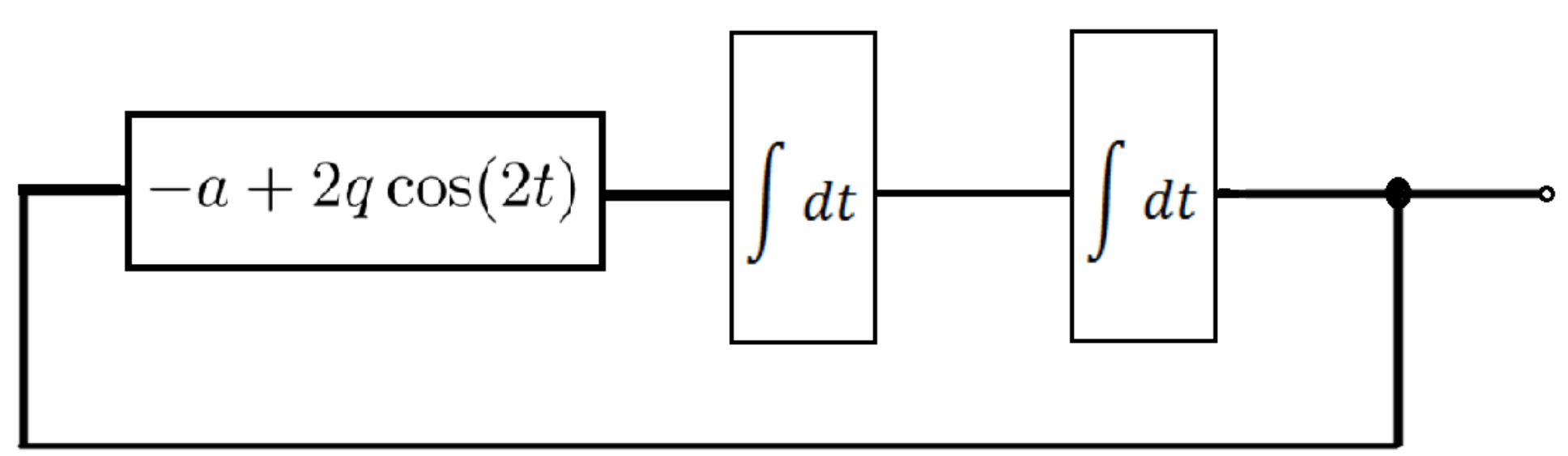}
\caption{Schematic drawing of an analog circuit for 
the integration of the Mathieu equation.}
\label{fig:circuitMathieu}
\end{figure}

Let us consider the Mathieu equation \cite{mclachlan}
\begin{equation}
{\ddot y}(t) + [ a - 2  q  \cos  (\omega t)] y(t) = 0,
\label{eq:mathieu}
\end{equation}
with $a$ and $q$ real constants \cite{note_mathieu}. 
If we define $p\equiv \dot{y}$, we 
can write the Mathieu equation as
\begin{equation}
\left [
\begin{array}{c}
{\dot y}(t) \\
{\dot p}(t)
\end{array}
\right ] = 
\left [
\begin{array}{cc}
0 & 1 \\
-  [ a - 2  q  \cos  (2  t)]   & 0
\end{array}
\right ]
\left [
\begin{array}{c}
 y(t) \\
 p(t)
\end{array}
\right ].
\end{equation}
Similarities and differences
between this equation and the Schr\"odinger equation 
(\ref{eq:schrodingerRabi}) for the semiclassical Rabi model
are self-evident. This point can also be visualized by comparing
the analog circuit of Fig.~\ref{fig:circuitMathieu} 
for the Mathieu equation with the circuit of 
Fig~\ref{fig:circuitRabi} for the semiclassical Rabi model. 

\begin{figure}[t]
\includegraphics[angle=0.0, width=8.5cm]{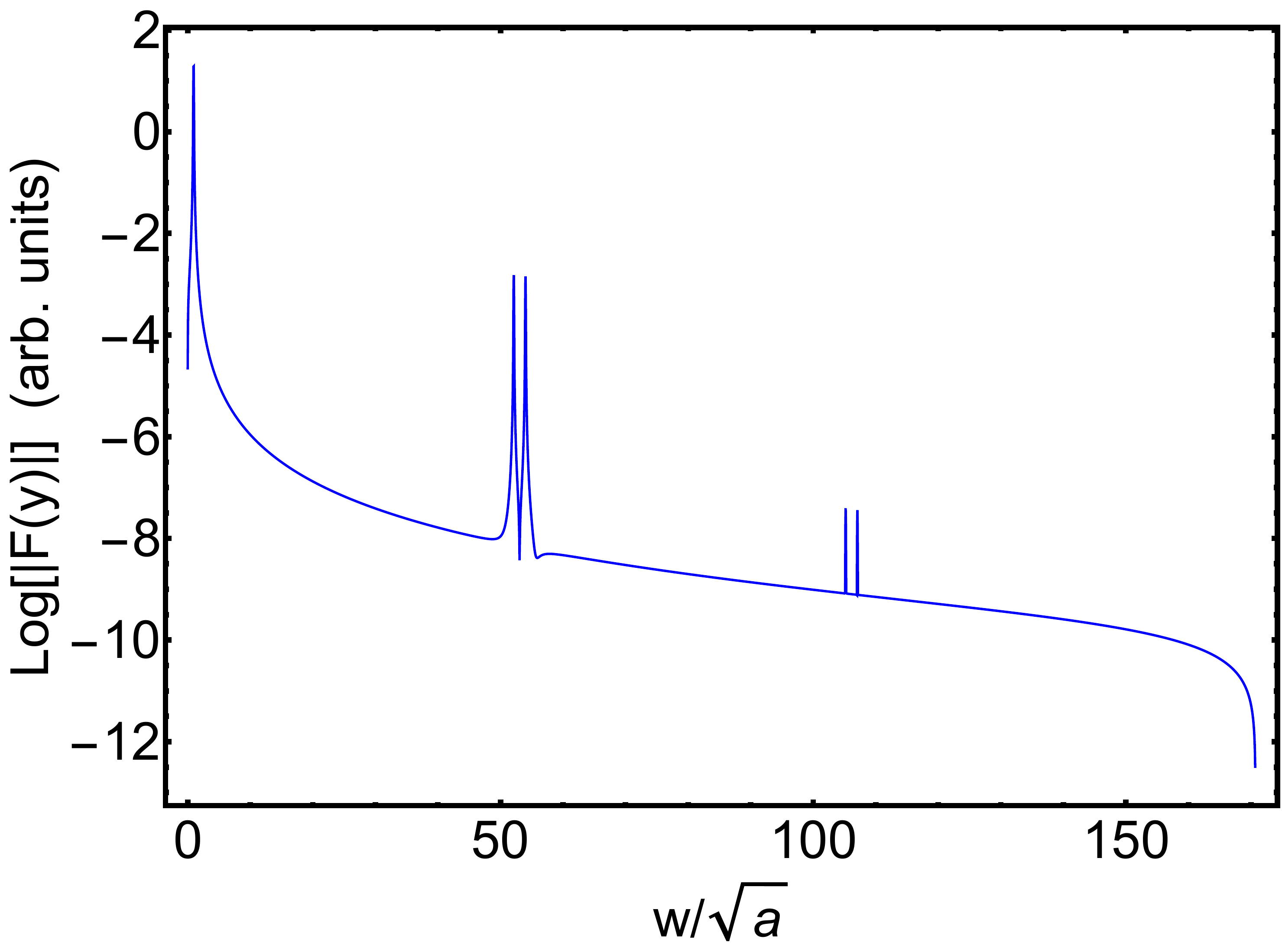}
\caption{Fourier transform 
$[F{y}](w)$ of the solution $y(t)$ of the Mathieu 
equation (\ref{eq:mathieu}) with initial
conditions $y(t=0)=1$, $p(t=0)=0$,
integrated up to $t=200$.
Parameter values: $a=0.5$, $q=0.1$, and $\omega=40$.}
\label{fig:doubletsMathieu}
\end{figure}

As for the semiclassical Rabi model, there are doublets in the 
Fourier spectrum, see Fig.~\ref{fig:doubletsMathieu}. 
Such doublets are generated by time-dependent terms, that is,
in the Rabi model the terms beyond RWA, and in the Mathieu equation
the term $2q\cos(\omega t)$, which corresponds to a frequency modulation. 
For the Rabi model the evolution is unitary, 
while this is not the case for the Mathieu equation. 
However, this lack of unitarity does not affect in any way the reason 
why doublets are present. 
On the other hand, we shall limit ourselves 
to the stable region with small values of the parameters $a$ and $q$,
since for large values of these parameters there are strong nonlinearities 
and a more complicated treatment is needed.
We can see from Fig~\ref{fig:doubletsMathieu} that in the Fourier
transform $[F{y}](w)$ of $y(t)$ there is a single peak at 
$w=\sqrt{a}$ and doublets at frequencies 
$n\omega \pm \sqrt{a}$, with
$n=1,2,3,...$. Similarly to the semiclassical Rabi model, 
this non-trivial structure may be explained on the basis of the 
Picard expansion. 

\begin{thebibliography}{100}

\bibitem{bourassa}
J. Bourassa, J. M. Gambetta, A. A. Abdumalikov, Jr., O. Astafiev,
Y. Nakamura, and A. Blais,
Phys. Rev. A \textbf{80}, 032109 (2009).

\bibitem{gross}
T. Niemczyk, F. Deppe, H. Huebl, E. Menzel, F. Hocke, M. J. Schwarz,
J. J. Garc\'{\i}a-Ripoll, D. Zueco, T. H\"{u}mmer, E. Solano,
A. Marx, and R. Gross, Nature Phys. \textbf{6}, 772 (2010).

\bibitem{mooij}
P. Forn-D\'{\i}az,
J. Lisenfeld, D. Marcos, J. J. Garc\'{\i}a-Ripoll, E. Solano,
C. J. P. M. Harmans, and J. E. Mooij,
Phys. Rev. Lett. \textbf{105}, 237001 (2010).

\bibitem{lupascu}
P. Forn-D\'{\i}az, J. J. Garc\'{\i}a-Ripoll, B. Peropadre,
J.-L. Orgiazzi, M. A. Yurtalan,	R. Belyansky, C. M. Wilson, and A. Lupascu,
Nature Phys. \textbf{13}, 39 (2017).

\bibitem{semba}
F. Yoshihara, T. Fuse, S. Ashhab, K. Kakuyanagi, S. Saito, and K. Semba,	
Nature Phys. \textbf{13}, 44 (2017).

\bibitem{moore}
G. T. Moore, J. Math. Phys. (N.Y.) \textbf{11}, 2679 (1970).

\bibitem{dodonov}
V. V. Dodonov,  Phys. Scripta \textbf{82}, 038105 (2010).

\bibitem{noriRMP}
P. D. Nation, J. R. Johansson, M. P. Blencowe, and F. Nori,
Rev. Mod. Phys. \textbf{84}, 1 (2012).

\bibitem{jaskula}
J.-C. Jaskula, G. B. Partridge, M. Bonneau, R. Lopes, J. Ruaudel, D. Boiron,
and C. I. Westbrook,
Phys. Rev. Lett. \textbf{109}, 220401 (2012).

\bibitem{koghee}
S. Koghee and M. Wouters,
Phys. Rev. Lett. \textbf{112}, 036406 (2014).

\bibitem{solano2014}
S. Felicetti, M. Sanz, L. Lamata, G. Romero, G. Johansson, P. Delsing,
and E. Solano,
Phys. Rev. Lett. \textbf{113}, 093602 (2014).

\bibitem{johansson2015}
C. Sab\'{\i}n, I. Fuentes, and G. Johansson,
Phys. Rev. A \textbf{92}, 012314 (2015).

\bibitem{adesso2015}
C. Sab\'{\i}n and `G. Adesso, 
Phys. Rev. A \textbf{92}, 042107 (2015).

\bibitem{savasta2015}
R. Stassi, S. De Liberato, L. Garziano, B. Spagnolo, and S. Savasta,
Phys. Rev. A \textbf{92}, 013830 (2015). 

\bibitem{exotic}
G. Benenti, S. Siccardi, and G. Strini,
Eur. Phys. J. D \textbf{68}, 139 (2014).

\bibitem{casimirqip}
G. Benenti, A. D'Arrigo, S. Siccardi, and G. Strini,
Phys. Rev. A \textbf{90}, 052313 (2014).

\bibitem{frigo}
G. Benenti and G. Strini, 
Phys. Rev. A \textbf{91}, 020502(R) (2015).

\bibitem{DCEoptimal}
F. Hoeb, F. Angaroni, J. Zoller, T. Calarco, G. Strini, S. Montangero, 
and G. Benenti, Phys. Rev. A \textbf{96}, 033851 (2017). 

\bibitem{norinature}
C. M. Wilson, G. Johansson, A. Pourkabirian, M. Simoen,
J. R. Johansson, T. Duty, F. Nori, and P. Delsing,
Nature (London) \textbf{479}, 376 (2011).

\bibitem{lahteenmaki}
P. L\"ahteenm\"aki, G. S. Paraoanu, J. Hassel, and P. J. Hakonen,
PNAS \textbf{110}, 4234 (2013).

\bibitem{nomoto}
K. Nomoto and R. Fukuda, Progr. Theor. Phys. \textbf{86}, 269 (1991).

\bibitem{micromaser}
P. Meystre and M. Sargent III,
\textit{Elements of quantum optics} (4th Ed.)
(Springer--Verlag, Berlin, 2007).

\bibitem{QRM}
For a recent collection of articles on semiclassical and 
quantum Rabi models, see D. Braak, Q.-H. Chen, M. T Batchelor, 
and E. Solano, 
J. Phys. A: Math. Theor. \textbf{49}, 300301 (2016).

\bibitem{antiDCE}
I. M. de Sousa and A. V. Dodonov,
J. Phys. A: Math. Theor. \textbf{48}, 245302 (2015);
D. S. Veloso and A. V. Dodonov, 
J. Phys. B: At. Mol. Opt. Phys. \textbf{48}, 165503 (2015). 

\bibitem{motazedifard}
A. Motazedifard, M. H. Naderi, and R. Roknizadeh,
J. Opt. Soc. Am. B \textbf{32}, 1555 (2015).

\bibitem{Blanes2009}
S. Blanes, F. Casas, J.~A. Oteo, and J. Ros,
Phys. Rep. \textbf{470}, 151 (2009).

\bibitem{moan}
P. C. Moan and J. Niesen,
J. Found. Comp. Math. \textbf{8}, 291 (2008).

\bibitem{casas2007}
F. Casas, 
J. Phys. A \textbf{40}, 15001 (2007).

\bibitem{mclachlan}
N. W. McLachlan,
\textit{Theory and Applications of Mathieu Functions}
(Oxford University Press, 1947).

\bibitem{note_mathieu}
We use a parametrization of the Mathieu equation which makes it
easier to compare with the semiclassical Rabi model. The standard 
Mathieu equation instead reads \cite{mclachlan}
\begin{equation}
{\ddot y}(t) + [ a - 2  q  \cos  (2 t)] y(t) = 0.
\end{equation}

\end{thebibliography}
\end{document}